\documentclass[a4paper,11pt]{article}
\usepackage{pos}

\usepackage{bm} 
\usepackage{color} 
\definecolor{darkgreen}{RGB}{0,128,0}
\definecolor{darkred}{RGB}{139,0,0}
\usepackage{slashed} 
\usepackage{hyperref}
\usepackage{ulem}
\newcommand{\beq}{\begin{equation}}
\newcommand{\beql}[1]{\begin{equation}\label{#1}}
\newcommand{\eeq}{\end{equation}}
\def\bal#1\gal{\begin{align}#1\end{align}}
\newcommand{\ball}[1]{\bal\label{#1}}
\newcommand{\eq}[1]{(\ref{#1})}
\newcommand{\fig}[1]{Fig.~\ref{#1}}

\renewcommand{\b}[1]{{\bm #1}}

\newcommand{\e}{\varepsilon}

\title{Photon radiation from quark-gluon plasma by means of chiral anomalies  without magnetic field}
 \ShortTitle{Photon radiation by means of chiral anomalies}

\author*[a]{Kirill Tuchin}

\affiliation[a]{Department of Physics and Astronomy, Iowa State University,\\
Ames, Iowa, 50011, USA}

\emailAdd{tuchink@gmail.com}

\abstract{We discuss a new channel of photon emission from the quark-gluon plasma, which opens up as photon acquires a tachyonic mass in the presence of the  CP-odd topological domains generated by the chiral anomaly. It allows photon radiation through the decay $q\to q\gamma$ and annihilation  $q\bar q\to \gamma$  processes closely related to the chiral Cherenkov radiation. Unlike previous proposals this mechanism does not require an external magnetic field. The differential photon emission rate per unit volume is computed and shown to be comparable to the rate of photon emission in conventional processes. The presentation is based on Refs.~\cite{Huang:2018hgk,Tuchin:2019jxd}.}

\FullConference{%
  HardProbes2020\\
  1-6 June 2020\\
  Austin, Texas}

\begin{document}
\maketitle

\section{Introduction}\label{sec:i}

Photon radiation in heavy-ion collisions has been throughly investigated over the last half century. Nevertheless the conventional approaches based on the perturbation theory are still struggling to give a comprehensive description of the photon spectra. This article reviews a novel mechanism of photon radiation via the chiral anomalies of the gauge theories. Actually, photon radiation  via the chiral anomaly in the presence of the intense magnetic field was investigated in \cite{Fukushima:2012fg,Mamo:2013jda,Mamo:2015xkw}. In contrast, the mechanism presented here does not require any magnetic field.

The quark-gluon plasma (QGP), is believed to contain the topological $CP$-odd domains created by the random sphaleron-mediated  transitions between different QCD vacua \cite{Kharzeev:2009fn,Zhitnitsky:2013hs}. The dispersion relation of the electromagnetic field interacting with these domains is \cite{Carroll:1989vb,Lehnert:2004hq,Tuchin:2014iua,Tuchin:2017vwb,Yamamoto:2015maz,Qiu:2016hzd}
\ball{a3}
\omega^2= \b k^2+ \omega^2_\text{pl}+m_A^2\,,
\gal
where $\b k$ is the photon momentum, $\omega_\text{pl}$ is the plasma frequency and  $m_A^2= -\lambda \sigma_\chi \omega$ with $\sigma_\chi$ being the chiral conductivity \cite{Fukushima:2008xe,Kharzeev:2009pj} (assumed to be a constant),  $\lambda=\pm 1$ are two photon circular polarizations. At high enough photon energies and plasma temperatures $\omega_\text{pl}$ is but a small correction compared to $m_A$ and will be neglected in the following sections.  The fact that photon acquires a mass in the chiral medium opens new photon production channels similar to the Cherenkov radiation \cite{Lehnert:2004hq,Tuchin:2018sqe,Tuchin:2018mte,Stewart:2019xjh}.

\section{Photon radiation rate}\label{sec:d}

Photon emission by means of the chiral Cherenkov radiation mechanism can proceed via two channels: (i) the decay channel  $q\to q\gamma$ and (ii) the annihilation channel $q \bar q\to  \gamma$. The total photon radiation rate is the sum of rates of these two processes.  

\subsection{Decay channel}

The scattering matrix element for photon radiation in the decay channel $q(p)\to q(p')+\gamma(k)$ is given by
$S_D=(2\pi)^4\delta^{(4)}(p'+k-p) i\mathcal {M}_D$ where 
\ball{c5}
i\mathcal {M}_D
=-ie Q \frac{\bar u_{\b p' s'}\slashed{\epsilon}^*_{\b k \lambda} u_{\b p s}}{\sqrt{8\e\e' \omega V^3}}\,.
\gal
The components of the 4-vectors are $p= (\e, \b p)$, $p'= (\e', \b p')$ and $k=(\omega, \b k)$, $Q$ is quark charge and $m= gT/\sqrt{3}$ its thermal mass \cite{Arnold:2001ms}. I retained the relativistic normalization factors $(2p^0 V)^{-1/2}$ for each of the three fields, where $V$ is the normalization volume. 
The rate of photon production per unit volume can be computed as
\ball{c9}
d\Gamma_D= 2N_c\frac{\delta(\omega+\e'-\e)}{16 (2\pi)^5\e\e'\omega}\sum_{\lambda s s'}|i\mathcal {M}_D|^2f(\e)[1-f(\e')]d^3k d^3p\,.
\gal
where $f(\e)$ is the quark equilibrium distribution function and the small chemical potentials of quarks is neglected.  Performing the summation over the transverse photon polarizations one obtains 
\ball{c12}
\sum_{ss'}|\mathcal{M}_D|^2=\frac{2}{x^2(1-x)}\left[q_\bot^2(2-2x+x^2)+m^2x^4\right]\,,
\gal
where $x= \omega/\e$ the fraction of the incident quark energy carried away by the photon and  $\b q_\bot= x\b p_\bot-\b k_\bot$. Neglecting $m$ one obtains
\ball{c21}
\omega\frac{d\Gamma_D}{d^3k}=2N_c\frac{ e^2Q^2}{8(2\pi)^4 } |m_A^2|\int_0^\infty d\xi \left[ \xi^2+(1-\xi)^2\right]f(\omega(1+\xi))\left[ 1-f(\omega\xi)\right]\,,
\gal
where only the polarization that gives $m_A^2<0$ contributes. The low and high energy regions of  the spectrum read 
\ball{c23}
\omega\frac{d\Gamma_D}{d^3k}=0.73\cdot 2N_c\frac{e^2Q^2}{8(2\pi)^4 } |m_A^2|
\left\{\begin{array}{ll}\frac{3\zeta(3)}{(\beta \omega)^3}\,, &  \omega\ll T \\
 \frac{1}{\beta\omega}e^{-\beta\omega}\,,&  \omega\gg T\,.\end{array}\right.
\gal
Taking into account that $m_A^2$ is proportional to $\omega$, one finds that at $\omega\ll T$, the photon  of spectrum scales as $1/\omega^2$. Thus the total photon rate $\Gamma_D$ is dominated by soft photons $\omega\ll T$ that produce the large logarithm $\ln (T/m)$. 

\subsection{Annihilation channel}

\begin{figure}[ht]
      \includegraphics[height=5cm]{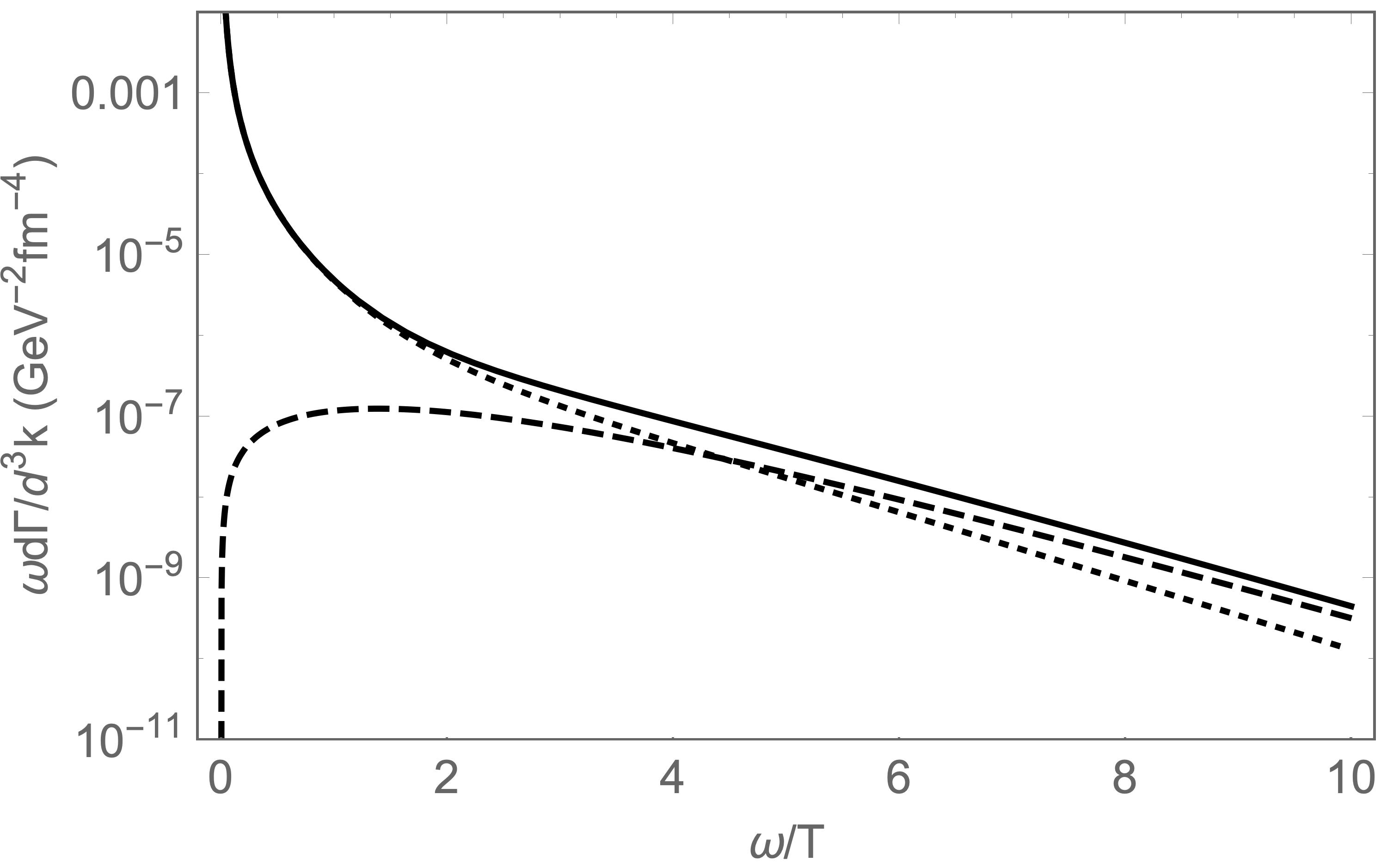} 
  \caption{Differential photon emission rate (solid line) and its two contributions from the decay (dashed line) and annihilation (dotted line) processes.  Plasma temperature $T=400$~MeV, chiral conductivity $\sigma_\chi = 1$~MeV  and $\sum_f Q_f^2 = 5/9$ (for the two lightest flavors). }
\label{fig1}
\end{figure}

The scattering matrix element for photon radiation in the annihilation channel $q(p)+\bar q(p_1)\to \gamma(k)$ is given by $S_A=(2\pi)^4\delta^{(4)}(p+p_1-k) i\mathcal {M}_A$ where 
\ball{d5}
i\mathcal {M}_A
=-ie Q \frac{\bar v_{\b p_1 s_1}\slashed{\epsilon}^*_{\b k \lambda} u_{\b p s}}{\sqrt{8\e\e_1 \omega V^3}}\,.
\gal
The corresponding  rate of photon production per unit volume reads
\ball{d9}
d\Gamma_A= \frac{dw_A}{VT}= N_c\frac{\delta(\omega-\e_1-\e)}{32 (2\pi)^5\e\e_1\omega}\sum_{\lambda s s_1}|i\mathcal {M}_A|^2f(\e)f(\e_1)d^3k d^3p\,.
\gal
Summation over the  photon polarizations and denoting by  $y= \e/\omega$ the energy fraction that the incident quark  contributed to the photon energy and $\b \ell_\bot= y\b k_\bot-\b p_\bot$ one derives 
\ball{d12}
\sum_{ss_1}|\mathcal{M}_A|^2=\frac{2}{y(1-y)}\left[\ell_\bot^2\left(y^2+(1-y)^2\right)+m^2\right]\,.
\gal
In the annihilation channel $m_A^2$ must be positive. Using this to compute the rate yields (in the chiral limit)
\ball{d22}
\omega\frac{d\Gamma_A}{d^3k}= N_c\frac{e^2Q^2 }{16(2\pi)^4 }|m_A^2|\int_0^1  dy f(y\omega)f((1-y)\omega)(2y^2-2y+1)\,.
\gal
 At low and high photon energy the spectrum reads 
\ball{d23}
\omega\frac{d\Gamma_A}{d^3k}=N_c\frac{e^2Q^2}{16(2\pi)^4 } |m_A^2|
\left\{\begin{array}{ll}\frac{1}{6}\,, &  \omega\ll T \\
 \frac{2}{3}e^{-\beta\omega}\,,&  \omega\gg T\,.\end{array}\right.
\gal
Comparing with \eq{c23} one can see that the decay channel dominates the low energy part of the spectrum, whereas the annihilation channel dominates the high energy tail, see \fig{fig1}.  It is remarkable that since the photon polarization in the two channels is opposite, the total photon spectrum has different polarization direction at low and high energies with respect to $T$. 

\section{Summary}\label{sec:f}

The main result of this work is Eqs.~\eq{c21} and \eq{d22} that represent the differential rates of  photon emission rate by means of the chiral Cherenkov radiation in the decay and annihilation channels. Their sum gives the total photon emission rate (per unit volume). The magnitude of this contribution to the total photon yield by QGP shown in \fig{fig1}  is comparable with the conventional contributions as one can glean from  Fig.~3 of \cite{Paquet:2015lta}.

The rate of photon production via the mechanism presented in this work is proportional to the chiral conductivity $\sigma_\chi$, which is in turn proportional to the axial chemical potential. This quantity is poorly known which is the source of the largest uncertainty in the photon rate. A more precise knowledge of $\sigma_\chi$ can be extracted from the measurements of the charge separation effect in relativistic heavy-ion collisions because it is generated by the anomalous electric current proportional to $\sigma_\chi$ \cite{Kharzeev:2007tn}.

\acknowledgments
This work was supported in part by the U.S. Department of Energy under Grant No.\ DE-FG02-87ER40371.


\begin{thebibliography}{99}
%
\bibitem{Tuchin:2019jxd}
K.~Tuchin,
Phys. Rev. C \textbf{99}, no.6, 064907 (2019).

\bibitem{Huang:2018hgk} 
  X.~G.~Huang and K.~Tuchin,
  Phys.\ Rev.\ Lett.\  {\bf 121}, no. 18, 182301 (2018).

\bibitem{Fukushima:2012fg} 
  K.~Fukushima and K.~Mameda,
  Phys.\ Rev.\ D {\bf 86}, 071501 (2012).
  

\bibitem{Mamo:2013jda} 
  K.~A.~Mamo and H.~U.~Yee,
  Phys.\ Rev.\ D {\bf 88}, no. 11, 114029 (2013).

\bibitem{Mamo:2015xkw} 
  K.~A.~Mamo and H.~U.~Yee,
  Phys.\ Rev.\ D {\bf 93}, no. 6, 065053 (2016).


\bibitem{Kharzeev:2009fn} 
  D.~E.~Kharzeev,
  Annals Phys.\  {\bf 325}, 205 (2010).
  [arXiv:0911.3715 [hep-ph]].
  
\bibitem{Zhitnitsky:2013hs}
A.~R.~Zhitnitsky,
Annals Phys. \textbf{336}, 462-481 (2013).
 

\bibitem{Carroll:1989vb}
  S.~M.~Carroll, G.~B.~Field and R.~Jackiw,
  Phys.\ Rev.\ D {\bf 41}, 1231 (1990).

\bibitem{Lehnert:2004hq} 
  R.~Lehnert and R.~Potting,
  Phys.\ Rev.\ Lett.\  {\bf 93}, 110402 (2004).


\bibitem{Tuchin:2014iua}
  K.~Tuchin,
  Phys.\ Rev.\ C {\bf 91}, 064902 (2015).

\bibitem{Tuchin:2017vwb}
  K.~Tuchin,
  Nucl.\ Phys.\ A {\bf 969}, 1 (2018).

\bibitem{Yamamoto:2015maz}
  N.~Yamamoto,
  Phys.\ Rev.\ D {\bf 93}, 085036 (2016).



\bibitem{Qiu:2016hzd}
  Z.~Qiu, G.~Cao and X.~G.~Huang,
  Phys.\ Rev.\ D {\bf 95}, 036002 (2017).
  
\bibitem{Fukushima:2008xe}
  K.~Fukushima, D.~E.~Kharzeev and H.~J.~Warringa,
  Phys.\ Rev.\ D {\bf 78} (2008) 074033.
  

\bibitem{Kharzeev:2009pj} 
  D.~E.~Kharzeev and H.~J.~Warringa,
  Phys.\ Rev.\ D {\bf 80}, 034028 (2009).

  
\bibitem{Tuchin:2018sqe}
K.~Tuchin,
Phys. Lett. B \textbf{786}, 249-254 (2018).

    
\bibitem{Tuchin:2018mte} 
  K.~Tuchin,
  Phys.\ Rev.\ D {\bf 98}, no. 11, 114026 (2018).
  
\bibitem{Stewart:2019xjh}
E.~Stewart and K.~Tuchin,
Phys. Rev. Research. \textbf{1}, 023005 (2019).

\bibitem{Arnold:2001ms} 
  P.~B.~Arnold, G.~D.~Moore and L.~G.~Yaffe,
  JHEP {\bf 0112}, 009 (2001).
  
\bibitem{Paquet:2015lta} 
  J.~F.~Paquet, C.~Shen, G.~S.~Denicol, M.~Luzum, B.~Schenke, S.~Jeon and C.~Gale,
  Phys.\ Rev.\ C {\bf 93}, no. 4, 044906 (2016).

\bibitem{Kharzeev:2007tn}
  D.~Kharzeev and A.~Zhitnitsky,
  Nucl.\ Phys.\ A {\bf 797} (2007) 67.
  

\end{thebibliography}
\end{document}